\documentclass[10pt,letterpaper,twocolumn]{article} 

\usepackage{ol2}
\usepackage[draft]{hyperref}
\usepackage{amsmath}

\begin{document}

\twocolumn[ 

\title{Zeeman slowing of thulium atoms}


\author{  K.~Chebakov, A.~Sokolov, A.~Akimov, D.~Sukachev, S.\,Kanorsky, N.~Kolachevsky,$^*$ V.~Sorokin  }

\address{P.N. Lebedev Physics Institute, Leninsky prospekt 53,
Moscow, 119991 Russia\\
Moscow Institute of Physics and Technology, Institutskii per. 9, Dolgoprudny, 141700 Russia \\
$^*$Corresponding author: kolik@lebedev.ru}

\begin{abstract} We demonstrate laser slowing of a hot thulium atomic beam
using the nearly closed cycling transition
$4\textrm{f}^{13}6\textrm{s}^2(^2\textrm{F}^\circ)(J=7/2)\leftrightarrow4\textrm{f}^{12}(^3\textrm{H}_5)5\textrm{d}_{3/2}6\textrm{s}^2(J=9/2)$
at $410.6$\,nm. Atoms are decelerated to velocities around 25\,m/s
by a 40\,cm Zeeman slower. The flux of slowed atoms is evaluated
as  $10^7\,\textrm{s}^{-1}\textrm{cm}^{-2}$. The experiment
explicitly indicates the possibility of trapping Tm atoms in a
magneto-optical trap.
\end{abstract}


] 

\noindent  Cooling of any new atomic species opens new horizons
for spectroscopy, atomic and quantum physics, degenerate gas study
and some technical applications. Compared to buffer gas cooling
 demonstrated for a variety of atoms and molecules (see e.g. \cite{Hancox}),
  laser cooling is not universal,
 but gives more selectivity and control on atoms also typically allowing for lower temperatures. Depending on the
level structure and atomic properties as well as availability of
laser sources, an individual approach to each element is
necessary.

Lanthanides with their peculiar level structure are of special
interest for cooling and trapping \cite{Hancox,Ban2005,
Monroe2002}. Laser-cooled ytterbium with the closed 4f$^{14}$
shell is readily used in laboratories for degenerate gas studies
\cite{Fukuhara2007} and optical frequency metrology
\cite{Barber2008}. In 2006 erbium atoms with a much more complex
level structure were successfully laser cooled and trapped
\cite{McClelland2006}, sub-Doppler cooling was demonstrated
\cite{Berglund2007} and a new type of magneto-optical trap (MOT)
was reported \cite{Berglund2008}.

In 2007 we studied laser cooling transitions in atomic thulium
\cite{Kolachevsky2007}. Tm has only one stable bosonic isotope
($^{169}$Tm, the nuclear spin number is $I=1/2$) and a single
vacancy in the 4f shell. There are two components of the
ground-state fine splitting with the electronic momentum quantum
numbers of $ J = 7/2$ (the lower one) and $J = 5/2$ (the upper
one) separated by $2.6\times10^{14}$\,Hz.
 The magnetic dipole transition
between these levels  at 1.14\,$\mu$m is of particular interest
for high precision frequency measurements since it is highly
immune to external perturbations \cite{Hancox,Ishikawa1997} and
has a $Q$-factor of $2\times10^{14}$. This relativistic
transition is highly sensitive to an $\alpha$ variation (its
energy scales as $\alpha^2Ry$, $\alpha$ is the fine structure
constant, $Ry$ is the Rydberg constant) and can be used in the
laboratory search for the $\alpha$ drift \cite{Fischer2003,
Rosenband2008}.

Our previous study  \cite{Kolachevsky2007} showed that laser
cooling should be feasible using the strong transition at
$410.6$\,nm between the ground state
$4\textrm{f}^{13}6\textrm{s}^2(^2\textrm{F}^\circ)(J=7/2,\, F=4)$
and the excited state
$4\textrm{f}^{12}(^3\textrm{H}_5)5\textrm{d}_{3/2}6\textrm{s}^2(J=9/2,\,F=5)$
with a natural line width of $\gamma=10.5(2)$\,MHz ($F$ is the
total momentum quantum number). The transition is not completely
cycling since the upper level decays to 6 neighboring
opposite-parity levels. The calculations showed that the branching
ratio is at the $10^{-5}$ level which is small enough for
efficient deceleration of atoms. Similar to the case of Er
\cite{McClelland2006}, no need for a repumping laser is expected.
In this Letter we demonstrate one-dimensional laser cooling of a
thulium atomic beam with the help of a Zeeman slower.

\begin{figure}[b]
\centerline{
\includegraphics[width=0.45\textwidth]{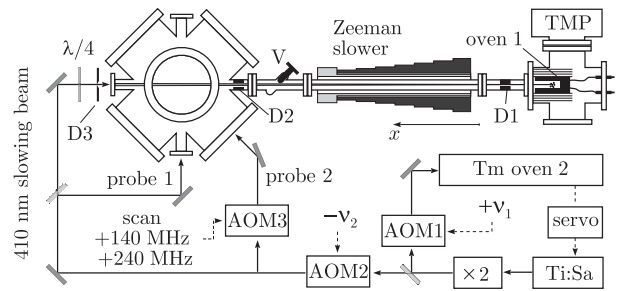}}
 \caption{\small Schematic of the setup. Here TMP is a turbo-molecular
 pump, D1 and D2 are diaphragms collimating the atomic beam, V is a blocking
 valve, and AOM is an acousto-optical modulator. Atoms may be excited either by the
 probe beam 1 at $90^\circ$ or by the probe beam 2 at $45^\circ$ in respect to their velocity.
 Luminescence photons coming from the center of the cross are detected by a photomultiplier orthogonal to the plane of the drawing (not
 shown).
 }\label{fig1}
\end{figure}

The experimental setup is shown in Fig.\,\ref{fig1}. Thulium metal
is sublimated in a home-made sapphire oven (oven 1) at a
temperature of 1100\,K measured by a platinum resistor. An atomic
beam with an intensity estimated as $10^9$ atoms per second is
formed by two diaphragms. The low-vacuum region containing the
oven is pumped by a 30\,l/s turbo-molecular pump and is separated
by the diaphragm D1 ($3$\,mm in diameter and $2$\,cm long) from
the rest of the vacuum chamber pumped by a 30\,l/s ion-getter
pump. Atoms pass the decelerating region and enter a 6-way-cross
vacuum chamber through the second diaphragm D2 with a diameter of
 $5$\,mm. The two vacuum volumes may be separated by a
 valve V.

\begin{figure}[b!]
\begin{center}
\includegraphics [width=0.45\textwidth]{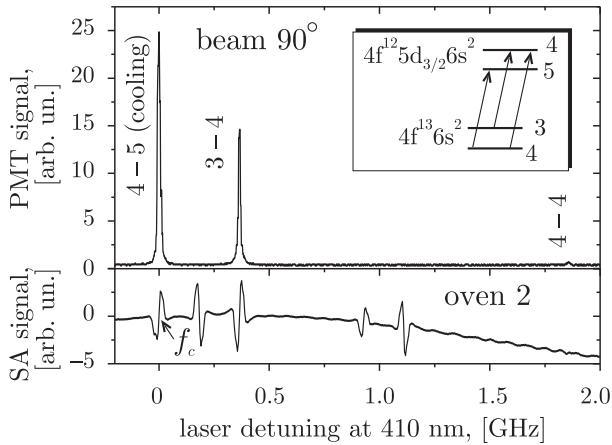}
\caption{ {\small Top: The spectrum of Tm atoms in the beam for
$90^\circ$ excitation (probe beam 1). Bottom: The saturation
absorption (SA) spectrum from the Tm oven. The servo loop is not
closed, laser frequency is scanned. The frequency axes are
corrected for AOM detunings to simplify identification. The inset
shows relevant Tm levels and allowed transitions between
$F$-sublevels.}
 }\label{fig3}
\end{center}
\end{figure}

The second harmonic of a Ti:sapphire laser is tuned to the
transition at 410.6\,nm. As a reference we use the saturation
absorption signal (Fig.\,\ref{fig3} (bottom))  from the oven 2 at
900\,K containing Tm chunks. The laser frequency may be either
freely scanned or locked to the saturation absorption signal
shifted by an acousto-optical modulator (AOM 1) working in the
$+1$ order at the frequency $\nu_1$. For locking we use the
cross-over resonance between the hyperfine $4\leftrightarrow5$ and
$3\leftrightarrow4$ transitions residing at $+180$\,MHz from the
cooling $4\leftrightarrow 5$ transition. The laser frequency $f_L$
is thus given by $f_L=f_c-\nu_1+180\,\textrm{MHz}$, where $f_c$ is
the frequency of the cooling transition.

The second AOM2 working in the $-1$ order (Fig.\,\ref{fig1}) at
the frequency $\nu_2$ formes the slowing light beam which is
detuned from the cooling transition by $\Delta
f_c=180\,\textrm{MHz}-\nu_1-\nu_2$ if the laser is locked to the
cross-over resonance.  The $g$-factors of the lower  and  the
upper level nearly coincide, the resulting sensitivity for
$\sigma$-polarized light is 1.46\,MHz/G. For the Zeeman slower
design taken from \cite{Barett1991} the red detuning $\Delta f_c$
of the slowing beam should be in the range from
$-130\,\textrm{MHz}$ to $-190\textrm{MHz}$ depending on the
desirable velocity of decelerated atoms.

 A part
of the beam is used as 90\,$^\circ$ probe (probe beam 1) which
crosses the atomic beam at the center of the chamber. The
luminescence photons are collected in the third direction and
focused on a photomultiplier (PMT). A typical Doppler-free
luminescence signal taken by scanning of the laser frequency (in
this case the laser is not locked) is shown in Fig.\,\ref{fig3}
(top). The observed line widths are consistent with the natural line
width $\gamma$. Imperfections of the probe beam 1 adjustments
result in a frequency uncertainty of the recorded lines of 5\,MHz.

\begin{figure}[t!]
\begin{center}
\includegraphics [width=0.45 \textwidth]{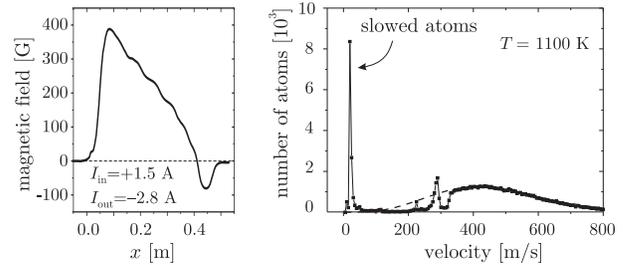}
\caption{\small{Left: Measured magnetic field on the axis of the
Zeeman slower. Right: Calculated longitudinal velocity
distribution at the exit of the slower (squares). The simulation
is made for $10^5$ atoms entering the slower with a Maxwellian
velocity distribution (dashed line) with a step of 5\,m/s.
Simulation parameters: the light field intensity equals
$20I_\textrm{sat}$, the detuning is $\Delta f_c=-140$\,MHz, the
currents are taken from the left part of the figure.}}\label{fig2}
\end{center}
\end{figure}

The 40-cm Zeeman slower consists of two coils of opposite
polarity \cite{Barett1991, McClelland2006}. The outer coil
consists of seven wire layers of different lengths which
effectively form 7 sections, while the inner one has two sections.
The outer coil (shown in black in Fig.\,\ref{fig1}) and the inner
coil (gray) are fed independently by currents $I_\textrm{out}$
and $I_\textrm{in}$ respectively. The measured axial magnetic
field distribution in the slower is shown in Fig.\,\ref{fig2}
(left). The currents mostly affect the number of slowed atoms
rather than their final velocity. The values shown in the figure
 are
experimentally optimized currents for the slower tuned to the
velocity 25\,m/s (see further).

Computer simulations of our Zeeman slower show that one can
decelerate 10-18\,\% of atoms from the atomic beam at the
temperature $T=1100$\,K to the velocity range of 10\,-\,40\,m/s.
An example  is shown in Fig.\,\ref{fig2} (right) with the magnetic
field distribution taken from Fig.\,\ref{fig2} (left). We assume a
homogeneous radial light field distribution with an intensity of
$20I_\textrm{sat}$, where $I_\textrm{sat}=6.1\,\textrm{mW/cm}^2$
is the saturation intensity. The detuning equals $\Delta
f_c=-140$\,MHz. Leaking of the atomic population to the ``dark''
states, losses on diaphragms and optical pumping from the $F=3$
sublevel are not taken into account. In this case about 10\,\% of
Tm atoms are decelerated to the velocities around 20-25\,m/s.

\begin{figure}[t!]
\begin{center}
\includegraphics [width=0.45\textwidth]{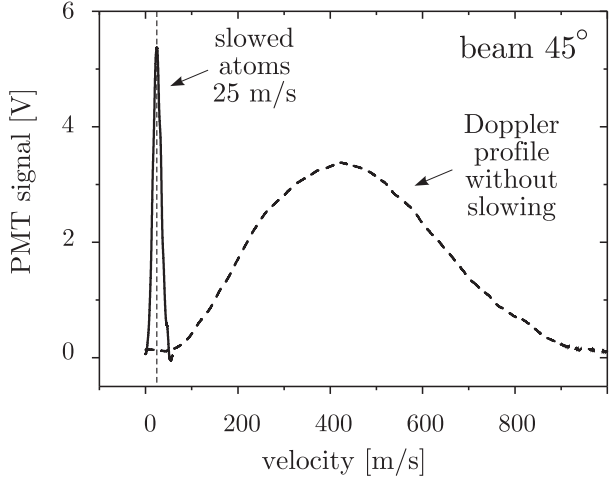}
\caption{\small{ The dashed  line is the luminescence spectrum of
Tm atoms for the $45^\circ$ excitation (probe beam 2) with the
Zeeman slower switched off. The servo loop is not closed, the laser
frequency is scanned.  The narrow peak residing at 25\,m/s (the
solid line) is the signal from slowed atoms taken at similar
experimental conditions with the Zeeman slower switched on. The
laser is locked, the AOM3 frequency is scanned.}}\label{fig4}
\end{center}
\end{figure}

The longitudinal velocity distribution of Tm atoms in the beam
$N_a(v_x)$ is probed by the laser probe beam 2 directed at
$45^\circ$ to the atomic beam (Fig.\,\ref{fig1}). The waist radius
(1/$e^2$) and the power of the probe beam are 1.6\,mm and
0.85\,mW correspondingly. To analyze the velocity distribution we
record the luminescence spectrum of the Tm beam by scanning the
laser frequency with the Zeeman slower switched off. The result of
this measurement is shown in  Fig.\,\ref{fig4}. To recover the
velocity distribution one should take into account the interaction
time with the laser beam $\propto1/v_x$. The PMT  was operating in
the current measuring regime to avoid non-linearities of the
photon counting. After the measurement the signal was calibrated.

The measured velocity distribution deviates from the thermal
distribution in a one-dimensional beam ($\propto
v_x^3\exp(-mv_x^2/2kT)$, where $m$ is the Tm atomic mass, $k$ is
the Boltzmann constant \cite{Greenland1985}). The attempt  to fit the
distribution from Fig.\,\ref{fig4} with the thermal one shows the
lack of slow atoms, which may result from our oven design (see
e.g.  \cite{Greenland1985,Scoles1988}).

To characterize the operation of the Zeeman slower the laser
frequency is locked to the cross-over resonance. The slowing laser
beam entering the vacuum chamber with a waist radius of 1.3\,mm is
additionally spatially filtered by a diaphragm D3 of 3\,mm in
diameter (Fig.\,\ref{fig1}) to reduce the scattered light. The
power of the slowing laser beam equals 8\,mW after D3 which
corresponds to an average intensity of $18I_\textrm{sat}$. The
probe beam frequency is scanned by AOM 3 working in the $+1$ order
which covers the  velocity range from $v_x=0$\,m/s to
$v_x=60$\,m/s. The recorded  data are corrected for the
frequency-dependent diffraction efficiency of the AOM.

The peak in Fig.\,\ref{fig4} demonstrates the operation of the
Zeeman slower optimized for $25$\,m/s. To remove the background,
two signals with the magnetic field switched on and off are
subtracted. Taking into account the measured PMT sensitivity of
$2\times10^7$\,photon\,s$^{-1}$\,V$^{-1}$, the photon collecting
efficiency of $7\times10^{-3}$ and the photon scattering rate in
the probe beam 2 of $2.5\times 10^7$\,s$^{-1}$ (averaged over the
probe 2 beam profile) we evaluate the flow of slow atoms through
the probe beam cross-section as $3\times10^6$\,s$^{-1}$. The
vertical size of the cross-section is the probe beam diameter
(3.2\,mm at $1/e^2$), while its horizontal size equals the atomic
beam diameter and is evaluated as 1\,cm. The latter results from
the diaphragms geometry and the angular spreading of slowed atoms
after the Zeeman slower. We evaluate the flux of the slowed atoms
as $\simeq10^7\,\textrm{s}^{-1}\textrm{cm}^{-2}$.

Comparing the count rates in the 25\,m/s peak with the total
velocity distribution taken at similar experimental conditions
(Fig.\,\ref{fig4}), the fraction of slowed atoms is evaluated as
1\,\% which is less than expected from the simulation
(Fig.\,\ref{fig2}). The difference is mainly explained by the
excessive angular spread of slow atoms at the exit of the Zeeman
slower which results in losses on diaphragm D2 and a change
of the atomic beam cross-section at the detection region. There
are other effects which may influence the ratio, e.g. the decay of
the upper cooling level
$4\textrm{f}^{12}(^3\textrm{H}_5)5\textrm{d}_{3/2}6\textrm{s}^2(J=9/2,\,F=5)$
to highly excited odd parity levels \cite{Kolachevsky2007} and
repumping from the other ground-state hyperfine component $F=3$ by
the cooling laser. Evaluations are consistent with the
experimental data. We demonstrated, that the velocity of  slowed
atoms can be varied in the range 20\,-\,40 m/s by changing the
detuning $\Delta f_c$ with corresponding optimization of the
magnetic field.

In conclusion, we have decelerated Tm atoms from a hot beam by
laser cooling at 410.6\,nm to the velocity range 20\,-\,40 m/s. The
beam of slow atoms of $1$\,cm in diameter has a flux of
$10^7$\,s$^{-1}\textrm{cm}^{-2}$ which should be enough for
capturing a cloud of $10^5$ Tm atoms in a MOT. No repumping laser
is necessary as in the case of laser cooling of Er
\cite{McClelland2006}.

This work has been supported by the RFBR grant 09-02-00649,
Presidential grants MK-1912.2008.2 and MD-887.2008.2, and the
Russian Science Support Foundation.

\end{document}